\documentclass[preprint2]{emulateapj}

\slugcomment{Received 2012 January 11; accepted 2012 July 12}

\shorttitle{Anomalous Low States in LMC X-3}
\shortauthors{Smale and Boyd}

\begin{document}


\title{Anomalous Low States and Long Term Variability
                  in the Black Hole Binary LMC X-3}


\author{Alan P. Smale and Patricia T. Boyd}
\affil{Astrophysics Science Division, NASA Goddard Space Flight Center, Greenbelt, MD 20771}
\email{alan.smale@nasa.gov, padi.boyd@nasa.gov}



\begin{abstract}

{\it Rossi X-ray Timing Explorer} observations of the black hole
binary LMC X-3 reveal an extended very low X-ray state lasting from
2003 December 13 until 2004 March 18, unprecedented both in terms of
its low luminosity ($>$15 times fainter than ever before seen in this
source) and long duration ($\sim$3 times longer than a typical
low/hard state excursion). During this event little to no source
variability is observed on timescales of $\sim$hours-weeks, and the
X-ray spectrum implies an upper limit of 1.2 $\times$
10$^{35}$ erg~s$^{-1}$. Five years later another extended low state
occurs, lasting from 2008 December 11 until 2009 June 17. This event
lasts nearly twice as long as the first, and while significant
variability is observed, the source remains reliably in the low/hard
spectral state for the $\sim$188 day duration.  These episodes share
some characteristics with the ``anomalous low states'' in the neutron
star binary Her X-1. The average period and amplitude of the variability
of LMC X-3 have different values between these episodes. We
characterize the long-term variability of LMC X-3 before and after the
two events using conventional and nonlinear time series analysis
methods, and show that, as is the case in Her X-1, the characteristic
amplitude of the variability is related to its characteristic
timescale. Furthermore, the relation is in the same direction in both
systems.  This suggests that a similar mechanism gives rise to the
long-term variability, which in the case of Her X-1 is reliably
modeled with a tilted, warped precessing accretion disk.

\end{abstract}





\section{Introduction}

LMC X-3 is a bright (up to 3$\times$10$^{38}$ erg~s$^{-1}$) black hole candidate
(BHC) in the Large Magellanic Cloud that is unique among
persistent BHCs in its high variability on timescales from days to years. The
mechanism governing this variability, and even the method by which the compact
object accretes mass from its companion, remains highly uncertain after decades of
observations and analysis. It is typically observed in the high/soft state, with
an X-ray spectrum qualitatively similar to that of other BHCs in the soft state: 
an ``ultrasoft'' component and a hard X-ray tail (e.g.\ \citet{ct96}).
Occasional, brief low/hard states show canonical spectral and timing behavior,
with a simple power law spectrum with photon index $\Gamma$=1.69$\pm$0.02,
a source luminosity at 50 kpc of $(5-16)\times 10^{36}$ erg~s$^{-1}$ (2-10 keV), and
strong broadband (0.01-100 Hz) time variability, with fractional rms amplitude of
40$\pm$4\% and a quasi-periodic oscillation peak at 0.46$\pm$0.02 Hz with rms
amplitude $\sim$14\% \citep{boyd00}.

The optical counterpart shows a large velocity range with
semi-amplitude K=235 km~s$^{-1}$ through its 1.7-day orbit. The lack of
eclipses implies an orbital inclination $<70^o$, and a compact object
mass of $\sim$7M$_\sun$ \citep{cowley83}. Historical UV and optical spectra
were used to constrain the additional variable flux, presumably from
the accretion disk, suggesting that at some orientations the disk is
responsible for nearly half of the UV/optical flux from the system \citep{cowley94}.
IUE and FOS spectra from 1992 show LMC X-3 to
have a UV line spectrum containing only weak emission features,
unusual for the presumed B3V mass donor star. The data show that as
LMC X-3 becomes brighter in the UV it also becomes bluer. Similarly,
the emission line strengths increase with UV flux. \citet{cowley91}
argue that this behavior is consistent with the accretion disk being
seen at varying observing angles, at some times obscuring and at other
times revealing the hot central part as it precesses. 
However, analysis of UV continuum and line spectra from {
\it IUE} and {\it HST}/FOS \citep{cowley94} timed to sample the long
term variability cycle showed that the UV and optical luminosity never
reached the expected maximum. The authors speculate than LMC X-3 may
have undergone an anomalous low state, however insufficient data were
available to test this hypothesis.

The X-ray contribution to the UV/optical spectrum complicates a
determination of the companion star's spectral type. \citet{soria01}
analyzed {\it XMM-Newton} Optical Monitor observations 
taken on 2000 April 19, when the source was near a
low/hard state \citep{boyd04}.
Fitting their results to evolutionary models for the companion,
Soria et al derived a  spectral type of B5IV. High resolution
optical spectroscopy of the companion star, with the appropriate
corrections for X-ray heating, leads to a similar spectral type
estimate of B5V \citep{val07}. The mass
accretion mechanism is therefore most likely Roche lobe overflow, with
minimal contribution from its stellar wind. In a related analysis, \citet{wu01}
analyzed {\it XMM-Newton} RGS and EPIC data from 2000
February--June, during the transition into and out of a low/hard
state. Spectral fits to the RGS data indicate that the line of sight
column density $N_H$ is $<10^{21}$ cm$^{-2}$, inconsistent with the expectation
of a larger value if the companion had a strong stellar wind. In
addition, no obvious emission lines were observed, consistent with no
previous epochs of wind matter ejection. \citet{soria01} conclude
that LMC X-3 accretes via Roche lobe overflow from the companion, as
opposed to via a strong stellar wind.

LMC X-3 belongs to a class of binaries that show high amplitude X-ray
variability on time scales much longer than their orbital
periods. Reports of a $\sim$198 (or perhaps $\sim$99) day superorbital
periodicity in early data \citep{cowley91} were based on unevenly
spaced data sets with gaps. More recent continuous monitoring
initiated with {\it RXTE}'s All-Sky Monitor
(ASM) in 1995 has led to the realization that the source shows a much
more complex and less strictly periodic long-term variability
\citep{paul00, wilms01, boyd04}.

\citet{wilms01} analyzed {\it RXTE} ASM light curves along with
Proportional Counter Array (PCA) spectra and found that LMC X-3
exhibits recurrent transitions from the soft to the hard state. They
argued in favor of a wind-driven limit cycle driving the observed
variability, and against a large accretion disk self-shadowing due to
a warp. 

From the analysis of 6 years of optical $V$ and $B$ data, partially
overlapping with {\it RXTE} ASM monitoring, \citet{brock01}
found that the long-term optical and X-ray variability are
correlated, with the X-ray lagging the optical by about 5 to 10
days. They argue that this supports the mechanism of variable mass
accretion rate giving rise to the long-term variability and suggest an
accretion disk wind instability limit cycle to generate the observed
behavior.

Several other X-ray binaries show large amplitude variations in their
X-ray fluxes on timescales significantly longer than their orbital
periods. The canonical example is the eclipsing X-ray pulsar Her X-1,
in which the long-term X-ray light curve is dominated by a high
amplitude, double-peaked 35-day variation. The main features of this
periodicity can be successfully explained by the presence of a warped
or inclined accretion disk precessing with respect to the binary
orbital plane \citep{pried87}. Other systems that show nearly periodic
variations at least an order of magnitude longer than the orbital
period include the accreting X-ray pulsars SMC X-1 (50–-60 dy;
\citet{gruber84, woj00}, but see \citet{clark03}) and LMC X-4 (30.5
dy; \citet{lang81}, but see \citet{trow07}), as well as the
paradigmatic disk jet source SS 433 \citep{fabian79, abell79}. In
almost all cases, the observed behavior is found to be more complex
than the prediction of strict periodicity from a simple precessing
disk model. Even the canonical source, Her X-1, has been observed to
undergo prolonged anomalous low states (ALS's) in which the long-term X-ray
variation is only marginally detected, if at all \citep{still04}. More
detailed modeling of the radiation-driven warping instability
uncovered regions of physical parameter space that could give rise to
unstable, apparently chaotic disk precession with no stable long-term
period \citep{wijers99, ogilvie01}. \citet{foulkes10} performed a
sophisticated numerical investigation using smoothed particle
hydrodynamics (SPH) and found that warping occurs across a broader
range of parameters and remains more stable than previously thought.

The organization of this paper is as follows. In Section 2, we
describe the data set we used to study the long-term variability of
LMC X-3 as well as our time series analysis methods, and results. We
argue that the data support the interpretation that LMC X-3 underwent
two ALS's and that the long-term behavior, as characterized by average
excursion lengths and amplitudes, is measurably different between ALS's,
in essentially the same manner as we previously observed in Her
X-1. Section 3 discusses the implications of the similarities of
long-term variability in terms of a period-amplitude relation that
holds in both LMC X-3 and Her X-1. We also compare the
period-amplitude relation of both systems with various nonlinear
oscillator models. In Section 4 we present our conclusions on the
existence of a period-amplitude relation in the long-term light curves
of these two quite different X-ray binaries, and some potential
further directions of investigation.

\section{Data Analysis and Results}

\subsection{The Overall Light Curve}

In this paper we make use of extensive archival data from the All-Sky
Monitor (ASM) \citep{levine96} and Proportional Counter Array (PCA)
\citep{jahoda96}, on board the {\it Rossi X-ray Timing Explorer}
satellite ({\it RXTE}) \citep{bradt93}. The ASM is a set of three
scanning shadow cameras which cover the 1.5--12.0 keV energy range in
three broad bands. It scans most of the sky each 90-minute satellite
orbit, collecting X-ray intensity measurements in 90 second
dwells. Since the launch of {\it RXTE} in 1995 December the ASM has
generated an impressive archive of high-quality, evenly sampled time
series data on virtually all moderately bright X-ray sources. The ASM
light curves can be used to explore variability timescales from days
to decades. The PCA consists of five Xe Proportional Counting Units
(PCUs) with a combined collecting area of about 6500 cm$^2$, and
provides high time resolution and moderate spectral resolution in the
2-–60 keV energy range.

\begin{figure}
\includegraphics[scale=0.42, angle=90]{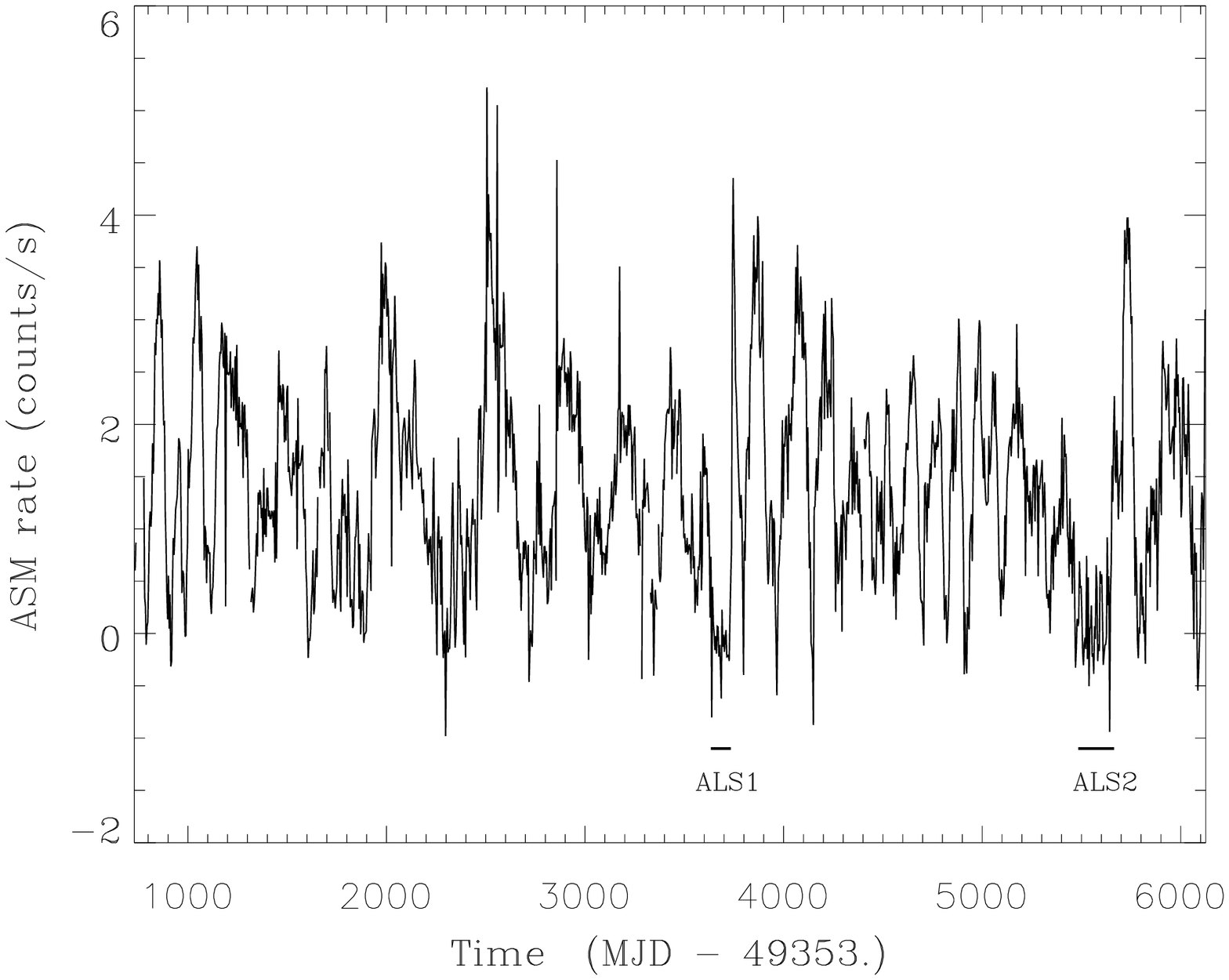}
\includegraphics[scale=0.37, angle=90]{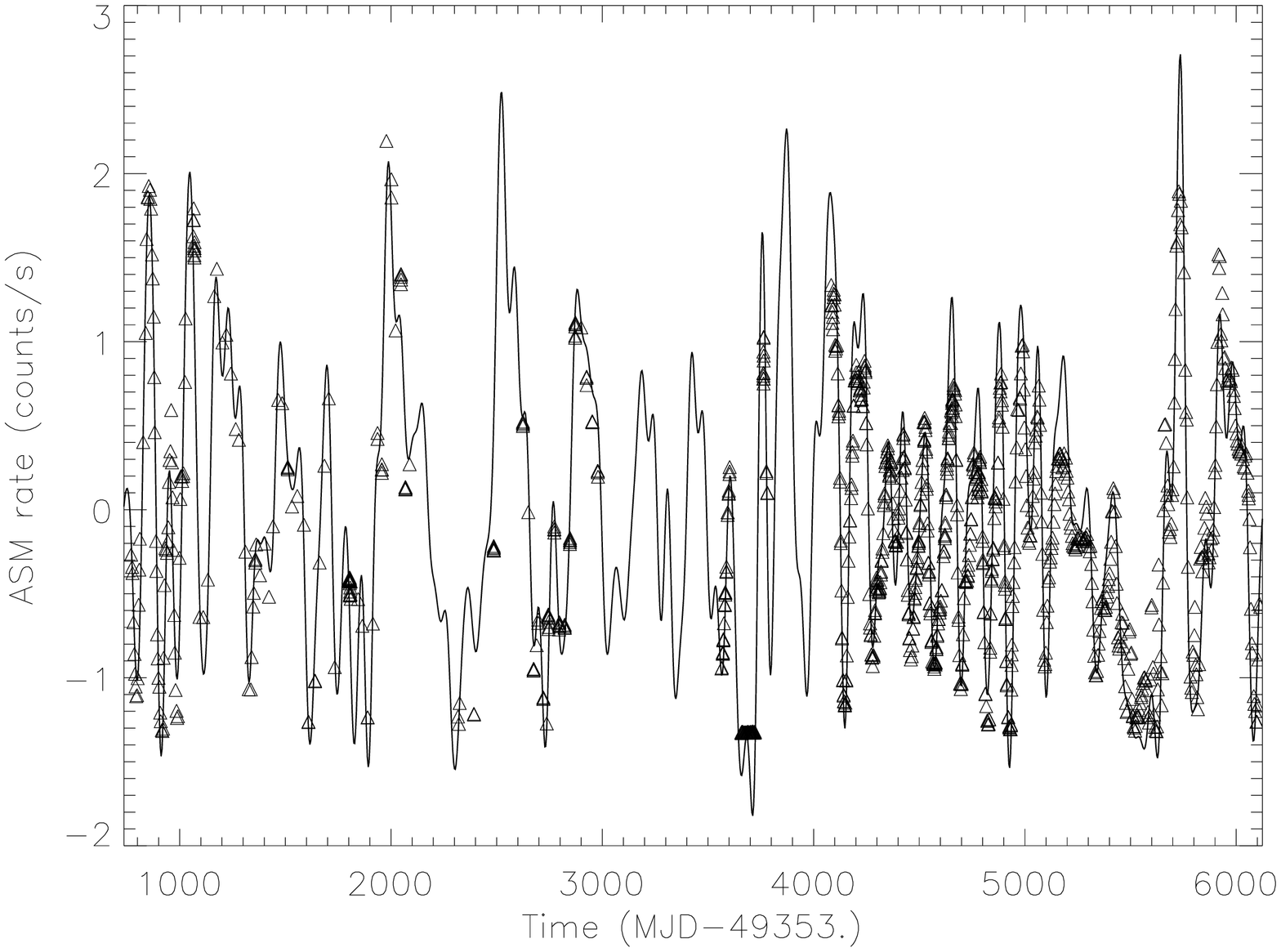}
\caption{
(a) The ASM light curve of LMC X-3 from 1996 through 2010, summed into
4-dy bins.  (b) The ASM light curve, smoothed as described in the text
(solid line), shown together with all available PCA Standard2 data
(triangles; one point per observation). Both data sets have been
mean-subtracted, and the PCA data has been scaled to have the same
maximum and minimum value as the ASM data. From this figure we
conclude that smoothed ASM light curve provides a faithful
representation of the overall source behavior, without adding spurious
artifacts.
\label{fig1}}
\end{figure}

The {\it RXTE} ASM light curve for LMC X-3 is remarkable in its
completeness, and is a virtually uninterrupted, evenly sampled time
series capturing the dramatic aperiodic variability from this unique
source (Figure 1a).  Typically, its X-ray flux is seen to vary
smoothly and continuously through a fairly fixed range, from fractions
of a count per second to just over 4 counts s$^{-1}$ (for comparison,
the Crab is approximately 74 counts s$^{-1}$). The timescale from one
maximum to the next varies from as short as two months to longer than
one year. In addition to this behavior, we note two unusual episodes
of remarkably low count rate, each lasting for several months (Figure
1a).

For our analysis of long-term variability, we started from the ASM
light curves available via the HEASARC. 
These contain data from each 90-second dwell, and include the
calibration and filtering criteria applied by the ASM instrument 
team\footnote{http://heasarc.gsfc.nasa.gov/docs/xte/ASM/asm\_events.html}.
We rebinned to a sample time of 4 days, a 
bin size chosen to improve the signal to noise ratio and minimize
the number of data gaps while preserving adequate sampling for the
timescales of interest. Having used the most recent ASM data calibration
available\footnote{http://heasarc.gsfc.nasa.gov/docs/xte/asm\_2010.html},
we restricted our analysis to data taken up to and
just beyond the 2010 recalibration to minimize variability due to
unmodeled calibration drift. Our dataset thus covers the 14.75 years
from 1996 January 5 through 2010 October 8.

For our analysis purposes, additional interpolation and filtering of
the ASM light curve was required. Since we require a truly evenly
spaced time series, the occasional sporadic gaps in the data were
filled with random data with the same mean and standard deviation as
the global dataset (only 18 data points out of 1,347 required this
treatment, representing 1.3\% of the total). A simple power
spectrum of the dataset reveals that the majority of the power is at
frequencies corresponding to periods of 50 days or more. The data were
thus smoothed using a Fourier low-pass filter, attenuating this
high-frequency noise. The resulting smoothed ASM light curve forms the
basis of Figure 1b, and is the principal light curve we utilize for
the bulk of the timing analysis described below.

The HEASARC public archive contains a wealth of observations of LMC
X-3 performed using the {\it RXTE} PCA. From launch to the end of 2010
there were a total of 1161 observations of LMC X-3, yielding a total
exposure time of 2.9 Msec. Since these were initiated in response to
19 successful Guest Observer pointed, monitoring, and Target of
Opportunity proposals, they comprise a heterogeneous and unevenly
spaced record. However, since early 2005 the time coverage of LMC X-3
has been outstanding, due to a succession of dedicated long-running
monitoring campaigns. The number of PCUs operating at any given time
is variable, and for the later datasets typically only two PCUs are
collecting data during monitoring observations. During every 
{\it RXTE} observation two standard PCA data modes are used, in addition
to user-specified modes chosen depending on the nature of the science
investigation. Standard2 mode collects all 129 spectral channels with
16 second time resolution, and due to its presence in all datasets
provides the core data we analyze here.

First, we use the complete PCA Standard2 data record to test the
possibility that our ASM light curve smoothing technique introduces
spurious variability, or fails to reproduce real long-term variability in the
source. In Figure 1b we show the smoothed ASM light curve for LMC X-3,
with the PCA data from all available {\it RXTE} pointed observations
superimposed. For each PCA observation we plot a single point, which
is the average, background-subtracted, Standard2 count rate per PCU
during the entire good observing time for that observation. It is
clear from Figure 1b that the gross morphology of the long term
variability, i.e., the occurrence times and flux values of the
extrema, as well as the overall character of the signal in between,
are very well preserved by the smoothed ASM light curve.

Inspection of Figure 1 shows that LMC X-3 has undergone a complex and
aperiodic evolution during almost fifteen years of continuous
monitoring by {\it RXTE}. Much of the time the flux is changing
slowly, smoothly and continuously between extremes. The
mean-subtracted time series appears stationary; there is no evidence
for the trends or modulations on $\sim$10-yr timescales which have
become apparent in other sources \citep{durant10}.

\subsection{Extended Low States}

Close examination of the ASM light curve reveals two long episodes of
low count rate. The unusual nature of these episodes can be emphasized
by plotting the logarithm of the ASM count rate versus time (Figure
2). For the purposes of this Figure we have added 2 to the
mean-subtracted ASM rate so that the signal is always positive
definite. The horizontal line indicates the level at which the flux is at least
1 standard deviation (1$\sigma$) below the mean value. 
When the data are plotted on this scale two things become
clear: (1) the flux value when the signal experiences a minimum does
not always go below even 1$\sigma$ of the mean value, and varies
considerably, and (2) there are two episodes where the source flux
stays very low for times much longer than the typical minimum. We
demarcate these episodes in Figure 2 using dashed vertical lines: the
extended low-state intervals correspond to {\it RXTE} mission days 3633-3729
and 5458-5646 (which in turn correspond to MJDs 52986-53082 and
54811-54999, and calendar dates 2003 December 13-2004 March 18,
and 2008 December 11-2009 June 17).

\begin{figure}
\begin{centering}
\includegraphics[scale=0.37, angle=90]{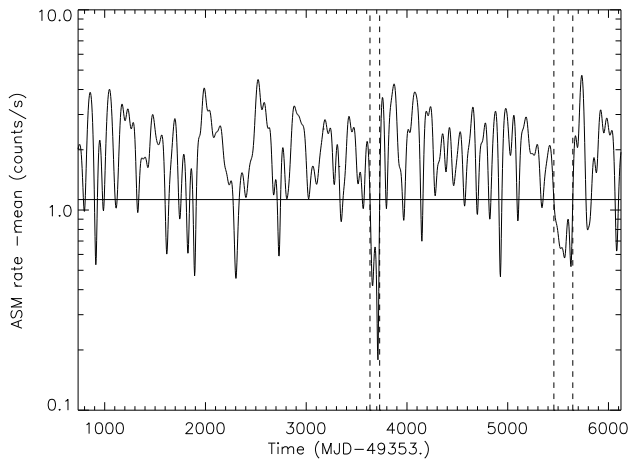}
\caption{ASM light curve for the same time period as Figure 1, but
with 2.0 added to the mean-subtracted count rate (see text), and the
flux now displayed on a logarithmic scale. The behavior of the
source when the flux is at least one standard deviation below the mean
(horizontal line) varies appreciably from one minimum to the
next. The two anomalously long episodes (dashed vertical lines) are
readily apparent in this representation.
\label{fig2}}
\end{centering}
\end{figure}

First, we compare these lows to the ensemble of lows in the system, to
see if they can indeed be considered anomalously long and low. To
identify the durations of episodes where the signal is truly low, we
locate times where the signal is 1$\sigma$ below the mean. For each
episode when the signal stays below this level, we use the local slope
of the signal to linearly interpolate to estimate the times when the
signal actually crosses this value, first in the downward and then in
the upward direction. We use these crossing times to characterize the
lengths of the lows. We find that there are 26 such low episodes,
including the two unusually long episodes identified above. The
shortest is 15 dy long. The longest lasts 188 dy. The average duration
of all 26 low episodes is 37.4$\pm$6.9 dy (where the error is defined as
$\sigma$/sqrt(n)). Excluding the two very long excursions and taking the
average of the remaining 24 episodes gives a mean duration of 28.7$\pm$2.3
dy. Using the same method of calculating crossing times, the duration
of the first long low state is found to be 96 dy (3.4 times longer
than the average and 6$\sigma$ outlier), while the second lasts 188 dy (6.6
times longer than the average and 14$\sigma$ outlier). Of the remaining
excursions, 21 are $<$35 days in duration. There are three low excursion
which last $>$35 days. They have lengths of 
39 days (centered at RXTE mission day 1618 = MJD 50971), 
52 days (centered at RXTE mission day 5802 = MJD 55155), and 
62 days (centered at RXTE mission day 2304 = MJD 51657).
PCA monitoring data covering the first two of these low excursions show
a continuous, smooth change from higher fluxes to the low point and back
up to higher flux levels, as seen in other typical low/hard excursions.  
There is insufficient PCA monitoring data through the third excursion,
however the raw 4-day ASM rate changes smoothly into and
out of this low, strongly suggesting that it too is typical.

We conclude that the two extended episodes of low flux we have identified
are indeed anomalously long. For reasons we discuss in detail below, we 
refer to them as Anomalous Low States 1 and 2 for the remainder of this Paper.

\

\subsection{Spectral and Timing Analysis of Low States in LMC X-3}

\subsubsection{Anomalous Low State 1 (ALS1)}

Fortuitously, we have excellent PCA coverage during ALS1. These
observations were originally scheduled under {\it RXTE} observation ID
(obsID) 80118-02 to be contemporaneous with an {\it INTEGRAL} study of
the LMC \citep{gotz06} supplemented with monitoring observations
performed under the second part of obsID 80103-01 to search for
hysteresis in state transitions, a study eventually completed using
different data obtained in 2005 March \citep{smith07}. To the best of
our knowledge, the PCA data we present in this paper have not been
analyzed or published previously.

As shown in Figure 3, PCA data using various combinations of PCUs were
obtained roughly every 3 days, for a total of 28 observations spread
over 67 separate obsIDs, yielding 152 ksec of good data which span
$\sim$64 days of the 96-day ALS (from 2004 January 7 through 2004
March 10). These PCA data were all obtained using Standard2 and Good
Xenon data modes.  To optimize the signal-to-noise ratio and enable
the most consistent measurements of spectral parameters and flux, we
limited our analysis to the 126.5 ksec of Standard2 data taken using
PCUs 2 and 3, Layer 1.  (The remaining good time includes data taken
with a different combination of PCUs.)

\begin{figure}
\includegraphics[scale=0.5, angle=90]{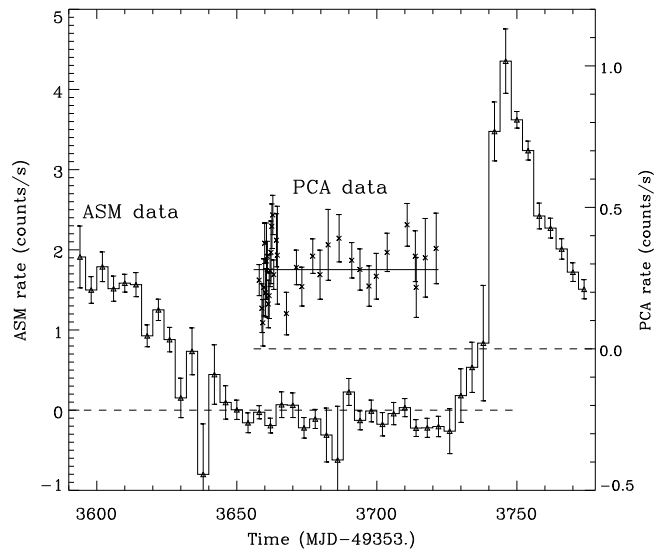}
\caption{
ASM and PCA light curves through the first anomalous low state. The
ASM data are in 4-dy bins, unsmoothed; the PCA data are plotted as a
single point per observation, with the measured count rate in the
3.0-25.0 keV band. Background subtraction for the PCA data does not
include a contribution for the LMC diffuse background (see text). 
\label{fig3}}
\end{figure}

The PCA data do not allow us to constrain the equivalent
hydrogen column density $N_H$, and we thus fix it at 3.8$\times$10$^{20}$ cm$^{-2}$ 
based on the accurate values obtained from radio measurements and from the O
I edge \citep{nowak01, page03, smith07}. Over
the energy range 3.0-25.0 keV, a good fit can then be obtained to the
summed background-subtracted PCA spectrum of ALS1 using a simple power
law model with photon index $\Gamma$=1.7$\pm$0.2, yielding a reduced
chi-squared slightly under 1. Neither the fit statistic nor the
residuals show a need to incorporate a second continuum component or a
line feature in the fit. For a distance of 50~kpc, the derived source
luminosity during ALS1 is L$_{x(2-10 keV)}$ = 4.2$\times$10$^{35}$ erg s$^{-1}$,
substantially lower than the typical luminosities derived for the
``canonical'' short low/hard states of LMC X-3 of (1.9-15.7)$\times$10$^{36}$ erg
s$^{-1}$ \citep{boyd00}.

However, this standard background subtraction does not take into
account the known large-scale diffuse emission from the LMC (see
e.g.\ \citet{points01}). Although this emission is difficult to model
and subtract out of our {\it RXTE} data, we can get a good indication
of its contribution by examining the {\it RXTE} PCA slews to and from
LMC X-3 for the ALS1 observations. We took two approaches to this,
firstly by fitting all the ALS1 on-source data using the nearby slew
data as ``background'', and secondly by carefully fitting spectra from
chosen slews. For the latter, we selected a series of slews for which
the Earth elevation angle was greater than 10$^o$ throughout, well
away from PCU breakdown events, and using only data from PCU2 for
which there are good slew data both onto and away from the
source. Both approaches give similar results: somewhere between 70\%
and 100\% of the measured flux from ``LMC X-3'' during ALS1 in fact
originates in the nearby diffuse LMC emission. Conservatively, this
would imply a maximum luminosity during ALS1 of
$\sim$1.2$\times$10$^{35}$ erg s$^{-1}$, $\sim$15 times fainter than
the minimum luminosity ever previously observed in a low/hard state,
representing 1.3$\times$10$^{-4}$ L$_{Edd}$, However, within the
errors, the data are consistent with the source having turned off
completely.

\subsubsection{Anomalous Low State 2 (ALS2)}

Our PCA coverage of ALS2 is also excellent, due to monitoring of LMC
X-3 carried out as part of the {\it RXTE} Core Observing Program. Under
obsIDs 93113-01 and 94113-01 data were obtained every 3--4 days
throughout the duration of ALS2 with good exposure times typically 2
ksec. Each of the individual spectra can be fit with a simple power
law model with index $\sim$1.7. Even the faintest of these spectra yields a
source detection at levels comfortably higher than that of the diffuse
LMC emission, and in fact significant variability continues to be
observed in LMC X-3 within ALS2. Constructing an aggregate spectrum
from the part of ALS2 where the flux levels are reliably the lowest
and least variable (for the 87 days lasting from 2009 January 29
through 2009 April 26, incorporating 47 ksec of good data) and then
fitting this spectrum with the equivalent hydrogen column density
fixed at 3.8$\times$10$^{20}$ cm$^{-2}$ as above, we achieve a good fit from 
3.0-25.0 keV with a power law index of $\Gamma$=1.71$\pm$0.02. The fit has a reduced
chi-squared of $\sim$1, and a derived 2-10 keV flux of 
4.62$\times$10$^{-11}$ erg cm$^{-2}$ s$^{-1}$, corresponding to a luminosity of 
1.38$\times$10$^{36}$ erg s$^{-1}$ (0.0015L$_{Edd}$)
at the distance of LMC X-3. This luminosity level is lower than, but
close to, those measured for the ``canonical'' low/hard states of LMC
X-3 studied by \citet{boyd00}.

In short, ALS2 is anomalous in duration, but not in the flux levels
reached, which are similar to those seen in the more customary,
shorter-duration low/hard states.

\subsection{Timing Analysis using Zero-Crossing Techniques}

In Figure 2, we provide dashed lines indicating where the anomalous
low states begin and end. Visual inspection of the data before and
after these ALS's suggests that the characteristic timescale and
amplitude of the long-term variability may be different. It appears
that the characteristic period is longer, and the amplitude higher
before the source enters ALS1 than after it exits. The system then
appears to favor a shorter-timescale, lower-amplitude variability until it enters
ALS2. While there are insufficient cycles of variability to claim a
new characteristic timescale after exiting ALS2, the $\sim$two available
cycles suggest a new longer timescale and higher amplitude than that
present between the two ALS's.

In an attempt to quantify and characterize this change in the
long-term variability, we consider the data outside of the ALS's as
three distinct groupings. We label the data before entering ALS1 on
{\it RXTE} day 3633 as ``early''. Data between the two ALS's (from {\it RXTE}
days 3729-5458) are labeled ``middle''. The data after exiting
ALS2 on day 5646 we label ``late''. For the purposes of characterizing
the long-term variability characteristics we do not consider the data
within the ALS's themselves.

Determining a timescale of interest from a nonperiodic signal requires
different analysis tools from a traditional analysis. Fourier analysis
is ideally suited to signals containing periodic, nearly sinusoidal
variability, but for astrophysical signals the variability often does
not fit this condition. Even a periodic signal as relatively simple as
that from the Crab pulsar results in a complex Fourier transform that
contains many harmonics as well as peaks at the fundamental
frequency. This is due to the shape of its pulse profile, which is
markedly steeper than a sine wave and has an interpulse offset from
phase 0.5 \citep{rots04}. For signals in which the period
and/or amplitude change with time, the use of a single Fourier
transform becomes even more problematic and can lead to
misidentifications of periodicities \citep{wilms01, paul00}. 

In cases where strict periodicity is not present in a
variable source, a short-time windowed Fourier transform (STWFT) has
been used to analyze signals where the variability is presumed to be
changing slowly with time. Even this assumption can potentially result
in misleading conclusions about the fine details of the time
evolution. In their reanalysis of the long-term evolution of SMC X-1,
\citet{trow07} used a minimum-fitting procedure to
determine the length of each individual superorbital cycle, They found
that the long-term ``period'' could change by as much as a factor of
two between adjacent cycles. Further, they found that the smoothing
imposed by the STWFT algorithm masked the measurable cycle-to-cycle
variability in the superorbital ``period'' for this source, which had
led some authors (e.g.\ \citet{clark03}) to infer a slowly and
smoothly varying long-term cycle perhaps due to the interaction of
multiple disk warps in the system.

\citet{still04} used a different method to measure individual cycle
lengths in the long-term period of Her X-1. To study variability in
the long-term period, they calculated an overall long-term ephemeris
from many cycles and then used this ephemeris to build an ``observed
minus computed'' diagram for all instances in which the source was
seen to be in the ``main-on'' high flux state. They found that this
more local, cycle-to-cycle variability analysis revealed a resetting
of the accretion disk clock between anomalous low states in the
system, whereas a STWFT analysis had appeared to indicate only a phase
shift. This and the previous example justify the use of a local
timescale analysis rather than Fourier or even STWFT analysis in order
to accurately extract the fine details of time variability in
non-periodic systems.

From the ASM light curve of LMC X-3 (Figure 1) it is clear that simply
setting a maximum or minimum count rate threshold and counting cycles
as occurring between either successive minima or maxima would result
in missing some cycles altogether and double-counting others, due to
the complex variability. Some maxima are much higher than others, some
minima much lower. However, the times when the mean-subtracted signal
changes value from negative to positive are quite smooth and well
behaved. Because of this, we adopt the method of measuring zero
crossings to characterize the cycle-to-cycle variability of LMC
X-3. Zero crossing analysis, a specific form of the more general level
crossing analysis, is a well accepted local time signal analysis
technique used in a wide variety of arenas, from speech processing to
oceanography to image analysis \citep{rab07, rych07, usac08, zhang10}.

Due to the significant spread in the value of the maxima of the LMC
X-3 light curve, and the fact that the minima are poorly defined due
to their intrinsically low count rate, the zero crossings are the most
easily localizable characteristic points in the signal, and identifying
these leads to the most straightforward estimate of the cycle
length from one oscillation to the next.

We find the time of each upward and downward zero crossing by
calculating the slope of the signal between adjacent points in time
where the sign changes from negative to positive or vice versa. The
slopes are then used to linearly interpolate to the time at which the
signal would have a value of zero. These are the zero crossings. The
``period'' for each long-term excursion is defined as the time
between two successive up-crossings (or two successive
down-crossings). The ``amplitude'' for each long-term excursion is
defined as the maximum minus the minimum flux value between two
successive crossings. For each cycle, two measurements of the period
and amplitude can be made, one using upcrossings and one using
downcrossings. Both approaches yield the same results within the
errors.

For a simply repeating, symmetric function such as a sine wave, the
period derived from the time between successive upcrossings would be
identical to that derived from successive downcrossings. Similarly,
the amplitude derived from the extrema between upcrossings would be
identical to that derived from the extrema between downcrossings. For
a time series such as LMC X-3, where the signal is not periodic and
the morphology of each flux excursion from one crossing to the next is
often far from symmetric, the individual periods derived from
individual upcrossings may vary significantly from those derived from
neighboring downcrossings (see Figure 4).

\begin{figure}
\includegraphics[scale=0.39, angle=90]{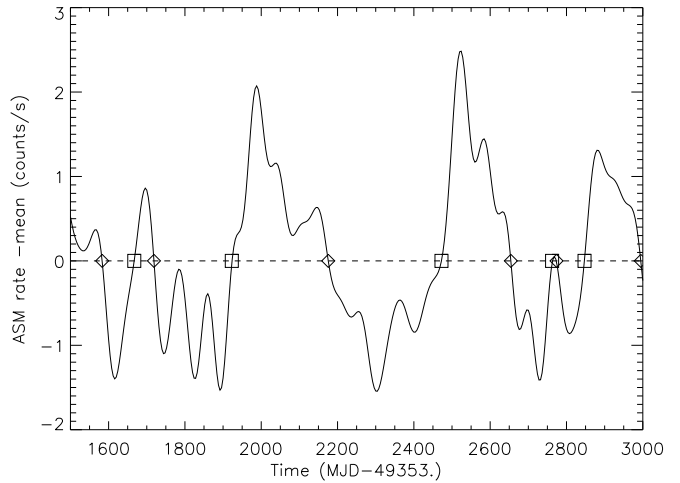}
\caption{ A segment of the mean-subtracted, smoothed ASM light curve
(solid line) is shown, with calculated times of zero upcrossings
(square symbols) and downcrossings (diamond symbols) indicated. For
irregular waves, in which the period and amplitude vary from one
wave to the next, the time between successive upcrossings or
downcrossings can be used to characterize the average period. Note
that, while differences are expected between the individual period
values obtained from the upcrossing versus downcrossing method, both
methods yield statistically similar values for period and amplitude
for irregular wave trains.
\label{fig4}}
\end{figure}

In the smoothed ASM lightcurve, LMC X-3 undergoes 14 upcrossings (and
so contains 13 upcrossing periods, where a period is defined as the
time between two successive upcrossings) and 15 downcrossings (14
downcrossing periods) in the early segment of the light curve. For the
middle segment, there are 14 upcrossing periods and 14 downcrossing
periods. For the late segment, there are three upcrossings (two
upcrossing periods) but only two downcrossings (and thus one single
downcrossing period).

We use these measured periods to compare the characteristics of the
long-term variability, as given by its average amplitude and period,
in the early, middle, and late data segments.  Figure 5 shows the
measured period and amplitude for each long-term flux excursion as
defined by upcrossings (squares) and downcrossings (diamonds). Table 1
lists the average period and amplitude derived from both the
upcrossing and downcrossing method, for the early, middle and late
data segments. While the individual values of period and amplitude
vary for the upcrossing versus downcrossing method, their average
values are within the errors (Figure 6). The errors are calculated
from the standard deviation of the period values from the mean,
divided by the square root of the number of cycles, N, in the data
segment. During the early data segment before ALS1, the average value
of the period is 213$\pm$36 days (upcrossing), and 204$\pm$35 days
(downcrossing). Between ALS1 and ALS2 (i.e., the middle segment), the
average period drops by around 80 days and the dispersion from the
mean falls three times lower than in the early segment to 119$\pm$10
days (upcrossing) and 118$\pm$10 days (downcrossing).  We note from
Figure 5 that the first few periods of the middle data segment are
higher than the middle average, and closer to the early average. 
However there is no excursion in the middle segment that reaches the
longer excursion lengths that are commonly seen in the early segment. Our
analysis suggests that the early segment and middle segment are
statistically different, however there is not enough data to
definitively conclude that the change in variability characteristics
is directly linked to the ALS's.

\begin{figure}
\includegraphics[scale=0.40, angle=90]{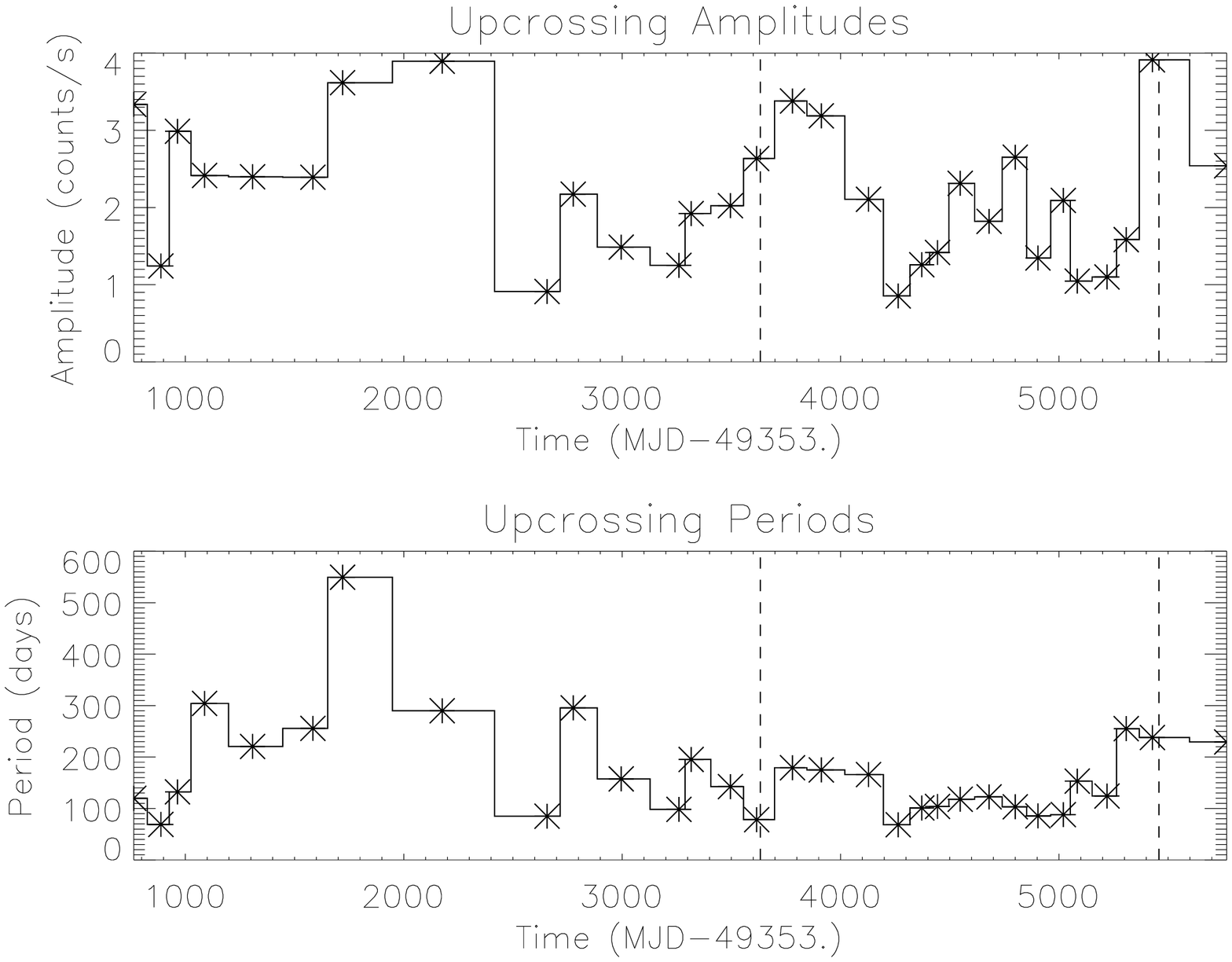}
\includegraphics[scale=0.40, angle=90]{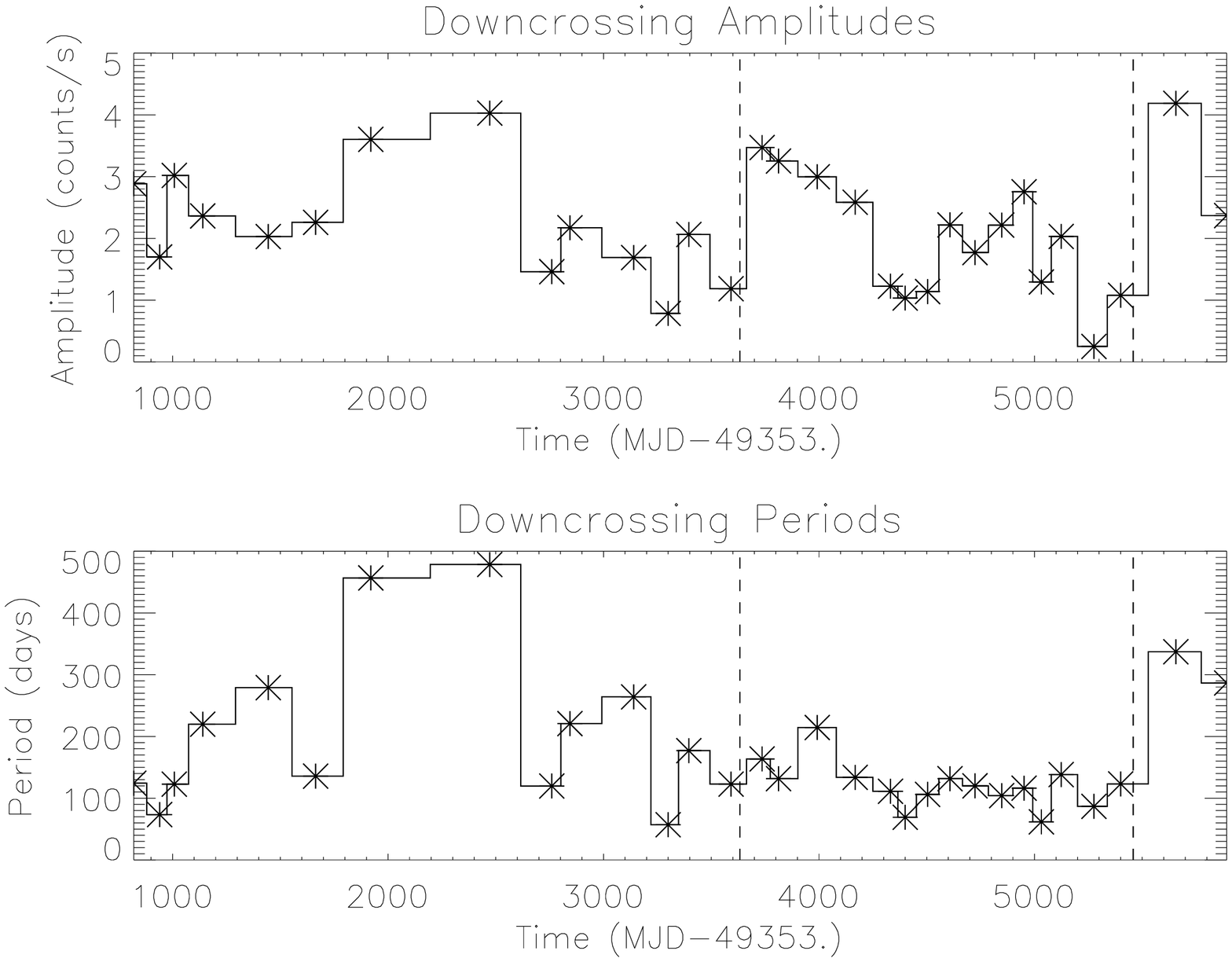}
\caption{
Wave-by-wave amplitudes (upper panel) and periods (lower panel)
derived from successive upcrossings and downcrossings. The
vertical dashed lines represent the mission start dates of ALS1 and
ALS2. The amplitudes, and in particular the periods, can be seen to
change their behavior after ALS1 (they cluster around a lower mean
value, and their dispersion is lower).
\label{fig5}}
\end{figure}

After exiting ALS2, there are only 2 upcrossing periods and a single
downcrossing period in the smoothed ASM light curve. These are listed
in the table for comparison. Although there are too few cycles for
statistical significance, the available data suggests that after ALS2,
LMC X-3 reenters a state that is more similar to the early segment,
with a longer period and higher amplitude, than the middle segment
which was characterized by a more tightly bound value of a lower
period, as well as a slightly lower amplitude. Note also that the
average period and amplitude appear to be monotonically related. This
also appears to be the case for individual cycles (Figure 7). As a
general rule, excursions with a longer period show a higher
amplitude. In particular, there are no occurrences of an individual
period longer than about 300 days whose amplitude is lower than 2
counts s$^{-1}$. Similarly an amplitude below about 1 count s$^{-1}$
has never been seen to occur with a period longer than about 100 days.

\section{Discussion}

\subsection{Anomalous Low States}

What is the physical state of the accretion disk in LMC X-3 during the
anomalous low states? Because the {\it RXTE} PCA is insensitive to
spectral components softer than $\sim$3 keV we cannot use these data
to provide robust constraints on a disk in its low-temperature,
low-flux states; a soft excess due to cool-disk emission from an
optically thin but geometrically thick accretion disk in the low/hard
state would lie beneath this energy cutoff. However, in the low/hard
states of GX 339-4, Cyg X-1, and Swift J1753.5$-$0127,
\citet{miller06} find evidence for a standard cool accretion disk
extending to the innermost stable circular orbit, down to L$_x$
$\sim$0.01L$_{Edd}$. For values of 0.001L$_{Edd}$ and below, models
predict that the entire inner disk may evaporate \citep{liu07,  taam08}. 
In addition, iron line studies of GX339-4 at low L$_x$ show that the
inner disk edge moves sharply outwards as L$_x$ decreases from
0.01L$_{Edd}$ to 0.001L$_{Edd}$ \citep{tomsick09}.  For LMC X-3 during
ALS1, where the upper limit on the derived total unabsorbed flux is
below 0.0005L$_{Edd}$ for an extended period, it seems likely that the
inner accretion disk has evaporated and is no longer present.

\subsection{The Long Term Variability}

From Figure 1, it appears that the amplitude of the long-term
variability is lower than average in the one or two cycles just prior
to each of the two ALS's. A similar trend occurs in Her X-1 just prior
to the start of its ALS's. This suggests that the entry into an ALS is not
instantaneous, but rather occurs in response to a slow decrease in the
accretion rate onto the central source. This then produces a decrease in
the flux from the central black hole, which could in principle bring it below
the critical value for which irradiation-driven disk warping can occur
\citep{wijers99, foulkes10}.

In Her X-1 \citet{still04} found that the amplitude of the long-term
variability directly scaled with the period of the $\sim$35-day
modulation, based on the analysis of three separate anomalous low
states observed in BATSE and ASM light curves between 1991 and
2004. Within the uncertainties, this period-amplitude relation appears
to also be present in LMC X-3. The behavior between the ALS's appears
to follow the same pattern, lending support to the idea that the
physical states of the accretion disk before, during and after an ALS
are similar in Her X-1 and LMC X-3, even though only Her X-1 is
consistent with stable disk precession. For Her X-1, \citet{still04}
interpret this as suggesting that the accretion disk becomes so
tightly wound that the effective radiation pressure felt by the disk
is reduced. Since the surface area on the disk incident to the compact
object decreases with increasing warp, the precession period is then
directly correlated with the flux of the compact object.
	
\begin{table*}
\begin{center}
\caption{Mean periods and amplitudes from zero crossings.\label{tbl-1}}

\begin{tabular}{llllll}
\tablewidth{0pt}
\tableline\tableline
 & Period & Period Error & Amplitude & Amp.Error & N \\
 & dy & dy & (counts s$^{-1}$) & (counts s$^{-1}$)  &  \\
\tableline
Early Up &       213 & 36 & 2.31 & 0.26 & 13 \\
Early Down & 204 & 35 & 2.23 & 0.24 & 14 \\
Middle Up &   119 & 10 & 1.94 & 0.22 & 14 \\
Middle Down & 118 & 10 & 1.85 & 0.23 & 14 \\
Late Up &        233 & 5 & 3.23 & 0.69 & 2 \\
Late Down &   287 & - & 2.36  &  -  & 1 \\
\tableline
\end{tabular}
\end{center}
\end{table*}

In general, nonlinear oscillators can be characterized by the relation
between the period and amplitude. In the linear regime, the period of
a simple oscillator such as a pendulum or spring is independent of the
amplitude, however this is not the case in the nonlinear regime. In
particular, soft spring oscillators (those that become less resilient
with elongation) have a period correlated with the amplitude of the
oscillation. For stiff springs, the reverse correlation is present. We
conjecture that the same period-amplitude relation is observed in LMC
X-3 and Her X-1 because the dynamics of the accretion disks in the two
systems are governed by a similar physical process. Furthermore, the
existence of a direct correlation between period and amplitude might
be predicted by some models of accretion disk dynamics, and not by
others. This means that observational results in such systems may give
us direct information about the underlying physics. For example, the
magnetorotational instability (MRI) is described mathematically in the
form of an oscillator, and has been shown to support chaotic solutions
(see e.g.\ \citet{winters03, um07}), and 
under some initial conditions to drive the development
of disk dynamos \citep{kapyla11}. This is closely related
to the mechanism that is thought responsible for the disk viscosity,
namely, the Balbus-Hawley MHD dynamo \citep{balbus91}. In this
scenario, when a disk possesses ionized hydrogen the MHD dynamo
operates, and localized regions of magnetic reconnection at the
surface of the disk cause sufficient electron heating to generate the
hard X-ray tail. Measuring observationally a well constrained value
for the period-amplitude dependence, coupled with spectral diagnostics
(such as a hard tail) may thus allow us to place some meaningful
physical limits on the presumed mechanisms responsible for both
driving and damping the disk oscillations, such as the electric charge
density of the disk material, and the strength of the disk magnetic
field.

\begin{figure}
\includegraphics[scale=0.4, angle=90]{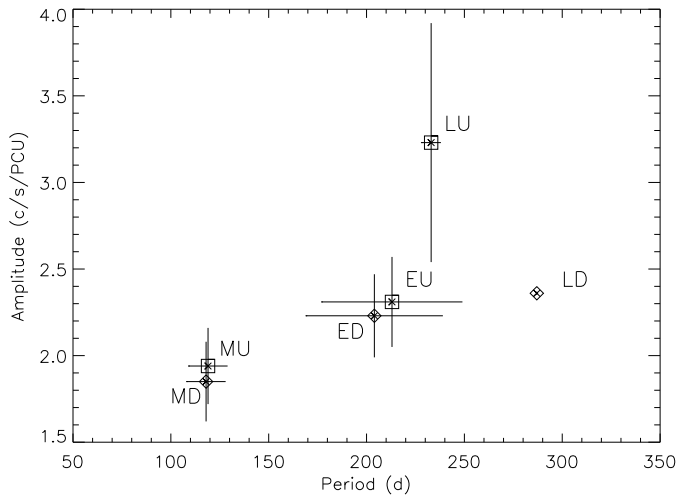}
\caption{
The average amplitude versus average period, derived from upcrossings
(labeled with U; marked with squares) and downcrossings (D; diamonds)
for early (E), middle (M) and late (L) data segments. For the late
segment there are only two upcrossing waves and a single downcrossing
wave.  Since the error bars represent the standard deviations from the
mean, the late downcrossing point has no error bar.
\label{fig6}}
\end{figure}

\begin{figure}
\includegraphics[scale=0.4, angle=90]{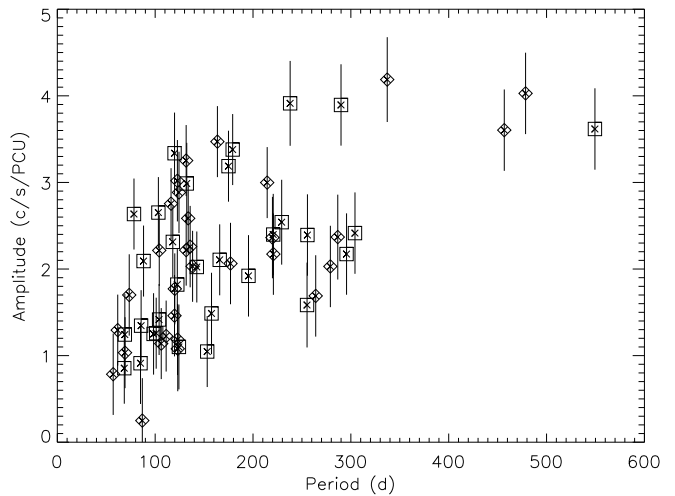}
\caption{
Wave amplitude versus period for each excursion, as derived from the
zero-crossing analysis. Data points marked with diamonds correspond to
downcrossings, those with squares to upcrossings. The period and
amplitude appear to be monotonically related. 
Excursions with longer periods have
higher amplitudes. There are no occurrences of an individual period
longer than about 300 days whose amplitude is lower than 2 counts
s$^{-1}$. Similarly an amplitude below about 1 count s$^{-1}$ has
never been seen to occur with a period longer than about 100 days.
\label{fig7}}
\end{figure}

Not only do the average amplitudes in LMC X-3 increase with increasing
period, but we have shown this relation seems to hold for individual
excursions as well (Figure 7). This could point to the possibility that the disk
is more similar to a so-called ``chaotic sea'' (used to characterize
ocean dynamics) than to a solid precessor, but hundreds of long-term
cycles would need to be observed before the distribution of cycle
heights and periods could be fit and compared to the expectations. A
similar cycle-to-cycle variability was noted in the distribution of
long-term variability excursions in the neutron star X-ray binary SMC
X-1 by \citet{trow07}.

\citet{foulkes10} present the results of 3D SPH simulations, using
realistic parameters for a wide range of X-ray sources. They find that
a generic flat disk is unstable to development of a warping mode, and
show that warped, precessing accretion disks occur more commonly in
X-ray binary systems than previously thought. In addition, their
simulations show that SMC X-1, Cyg X-2 and LMC X-3 - three systems
that all show varying long-term cycle lengths - all develop disks with
a similar maximum disk warp angle, disk tilt angle, and location of
maximum warp position. These three systems are all modeled by
\citet{foulkes10} as having very high compact object fluxes. To
describe the variable long-term cycle length in SMC X-1, Trowbridge et
al considered the possibility that the average radius at which
accreted material is injected into the disk could vary from one
long-term cycle to the next, as suggested by \citet{ogilvie01}.  In
fact, the simulations of \citet{foulkes10} support this suggestion by
revealing that, once precessing, the disk will continuously flex in
response to changes in the orientation of the Roche potential, which
results in changing the location of the accretion injection stream on
the disk.

\section{Conclusions}

The 15-year {\it RXTE} ASM monitoring data show that LMC X-3 has entered
into two distinct, extended low/hard states that share several
characteristics with the anomalous low states seen in Her X-1. The
characteristic amplitude and period of the long-term variability seen
in LMC X-3 is measurably different between ALS's. These differences
can be characterized statistically by calculating average excursion lengths and
amplitudes from a zero-crossing analysis applied to
the three segments of data. We find that the characteristic timescale is
monotonically related to the characteristic amplitude. A similar
relation is also present in Her X-1, and this relation is in the same
direction in each system. We interpret this as suggestive that the
accretion disk dynamics of LMC X-3 is on some level analogous to that
in Her X-1, even though the two sources contain a different type of
compact object at their centers. Whatever physical process is driving
the dynamics of the accretion disk resulting in the observed long-term
variability observed in Her X-1 may also be the dominant process
responsible for the long-term variability seen in LMC X-3. Analysis of
the variability in both systems suggests that this mechanism must be
sensitive to the initial conditions when the geometry governing the
dynamics is re-established, such as after ALS's. The existence of a
unifying mechanism giving rise to the long-term variability in two
very different systems - pulsar versus black hole - may help to
elucidate the process governing this still poorly understood
characteristic of many X-ray binaries, and tie it to a family of
nonlinear oscillators of a certain form.

\citet{wilms01} found that the specific long-term flux and
spectral variations of LMC X-3 could be well modelled by an accretion
disk wind-driven limit cycle driven by Compton heating, as proposed by
\citet{shields86}. In this scenario, under certain circumstances
that depend on the masses of the system parameters, the accretion rate
can become so high that as a result the compact object develops a wind
sufficiently strong to temporarily halt accretion. This loss of fuel
propagates forward in time, lowering the X-ray flux and damping the
wind, which then allows the accretion rate to again increase, the
recurrence happening on the viscous timescale. Since this model
includes a both damping and driving mechanisms, it deserves further
investigation to see if it could give rise to the behavior seen in LMC
X-3 and Her X-1.

Cycle-to-cycle variability is a hallmark of the long-term flux
evolution in a number of X-ray binary systems. However, characterizing
this variability is only possible with the existence of nearly
continuous, nearly evenly sampled lightcurve data on timescales longer
than the characteristic period of the variability under study. The
{\it RXTE} ASM has been the workhorse instrument in providing such data, and
we are now at the point where nonlinear time series analysis
techniques can be applied to real astrophysical data with variability
timescales of hundreds of days. With the quality of data in hand it
should soon be possible to extend this analysis to many other systems
that show high amplitude non-periodic long-term variability, to search
for other examples of such a period-amplitude relation, which could
point to an underlying physical mechanism that unlocks the secrets of
accretion disk dynamics.
 
\acknowledgments

This research has made use of data and software provided by the High
Energy Astrophysics Science Archive Research Center (HEASARC), which
is a service of the Astrophysics Science Division at NASA/GSFC and the
High Energy Astrophysics Division of the Smithsonian Astrophysical
Observatory. ASM light curves were provided by the {\it RXTE} ASM team at
MIT and NASA’s GSFC. We thank J. Heagy and J. Cannizzo for a careful
reading of the manuscript prior to submission.



{\it Facilities:} \facility{HEASARC}, \facility{RXTE}

\end{document}